\documentclass[%
 reprint,
 amsmath,amssymb,
 aps,
]{revtex4-2}

\usepackage{graphicx}
\usepackage{dcolumn}
\usepackage{bm}
\usepackage[breaklinks,colorlinks,linkcolor=blue,citecolor=red,urlcolor=blue]{hyperref} 
\usepackage{url}
\usepackage{color}
\usepackage{ulem}

\begin{document}

\preprint{APS/123-QED}

\title{The charge radii of calcium isotopes \textcolor{black}{within relativistic density functional theory}: nucleon’s finite-size and quadrupole shape fluctuation effects}

\author{H. H. Xie}
 \affiliation{College of Physics, Jilin University, Changchun 130012, China}

 \author{J. Li}
\email{jianli@jlu.edu.cn}
\affiliation{College of Physics, Jilin University, Changchun 130012, China}

\author{Y. L. Yang}
\affiliation{State Key Laboratory of Nuclear Physics and Technology,~School of Physics, Peking University, Beijing 100871, China}

\author{P. W. Zhao}
\email{pwzhao@pku.edu.cn}
\affiliation{State Key Laboratory of Nuclear Physics and Technology,~School of Physics, Peking University, Beijing 100871, China}

\date{\today}

\begin{abstract}
The anomaly in the charge radii of Ca isotopes has been puzzling for nuclear theory for decades.
We present the first self-consistent solution to this puzzle within the density functional theory without resorting to local parameter adjustment.
By taking into account both the intrinsic electromagnetic structure of nucleons and the zero-point motions of nuclear shape, which have been often neglected in previous studies, the similar charge radii of $^{40}$Ca and $^{48}$Ca as well as an inverted parabolic behavior between them are reproduced.
It is found that these effects also play crucial roles in the description of the isotonic shift between the charge radii of Sn and Cd isotopes.
\end{abstract}

\maketitle


Nuclear charge radius is one of the most fundamental properties of atomic nuclei.
It carries essential information about the saturation density of symmetric nuclear matter and provides a probe of nuclear force and nuclear many-body dynamics.
A significant amount of experimental data on charge radii has been collected across the nuclear chart~\cite{ANGELI201369,Campbell2016Prog.Part.Nucl.Phys.127}.
However, abundant experimental results leave puzzles to the theoretical side.
In particular, the anomaly in the charge radii along the Ca isotopic chain~\cite{neumann1976isotope, Palmer_1984,PhysRevLett.68.1679,garcia2016unexpectedly}, i.e., the almost equal values of charge radii in $^{40}$Ca and $^{48}$Ca and an inverted parabolic behavior between them, has been a longstanding challenge for nuclear theory in the past decades.

Nuclear density functional theory (DFT) is one of the most promising microscopic approaches for the study of charge radii throughout the nuclear chart.
While several state-of-the-art DFT calculations can provide quite accurate global descriptions of the experimental charge radii~\cite{PhysRevC.100.054306, PhysRevLett.110.032503, PhysRevC.98.044310, PhysRevC.82.035804, PhysRevC.89.054320, PhysRevC.103.054310, XIA20181}, they generally fail to describe the peculiar behavior of the charge radii in Ca isotopes.
This clearly implies that some important physics is missing in all these DFT calculations.
Although it has been \textcolor{black}{demonstrated} by Caurier et al.~\cite{CAURIER2001240} that the shell-model calculations could achieve a reproduction of charge radii for the Ca isotopic chain, the physical mechanism leading to peculiarly parabolic behavior is also study-worthy.

Within the framework of DFT, several theoretical efforts have been made to improve the description of charge radii for Ca isotopes.
The Fayans density functional Fy($\Delta r$)~\cite{FAYANS200049, PhysRevC.95.064328, miller2019proton} with an additional density-gradient term in the pairing interaction, as well as the relativistic density functional with a proposed correction term to charge radii associated with the Cooper pair~\cite{PhysRevC.102.024307}, can fairly reproduce the charge radii of Ca isotopes.
In both cases, however, the additional terms introduce parameters adjusted to fit the parabolic behavior of charge radii between $^{40}$Ca and $^{48}$Ca.

The effects of zero-point motions (ZPM) of the nuclear surface on ground-state charge densities and charge radii have been investigated in recent decades~\cite{Bohr1975, PhysRevC.28.355, BARRANCO198590, PhysRevLett.59.2724, FAYANS200049, Sil_2008}. 
In Ref.~\cite{PhysRevC.28.355}, the significant influence of ZPM on charge density for $^{40}$Ca and $^{208}$Pb has been investigated by means of a macroscopic treatment. 
Since then, it has been proposed 
that the inverted parabolic behavior could be correlated with the zero-point fluctuations with the assumption that the volume of the nuclear charge remains constant~\cite{BARRANCO198590}. 
Without local adjustment to charge radii of Ca isotopes, an alternative proposal is to consider the quadrupole deformation \textcolor{black}{(correlated with the effect of ZPM)} extracted from the $B(E2)$ values between the ground state and low-lying states~\cite{PhysRevC.106.L011304}, which is related to the increase in charge radius in the Bohr's collective model~\cite{Bohr1975}. 
Such corrections were found to significantly improve the description of charge radii in Ca isotopes with the SV-min density functional.
However, the correction was evaluated with the $B(E2)$ values either from experiments or shell-model calculations using effective interactions that have no connection to the original density functional. Therefore, a self-consistent interpretation for the anomalous behavior of charge radii in Ca isotopes is still in absence.

In this Letter, the long-standing puzzle of the charge radii between $^{40}$Ca and $^{48}$Ca is solved within the DFT framework in a self-consistent manner without resorting to additional parameter adjustments.
We propose the key role of both the intrinsic electromagnetic (EM) structure of nucleons and the zero-point motions of nuclear shape.
By taking into account their contribution, we show that the almost equal values of charge radii in $^{40}$Ca and $^{48}$Ca and an inverted parabolic behavior between them can be reproduced based on the relativistic density functional PC-PK1~\cite{PhysRevC.82.054319}.
It should be noted that the combination of mean-field-level density functional and beyond-mean-field shape-fluctuation corrections is not a new scheme proposed by the present work. In fact, such a scheme is quite common and successful in the description of many nuclear properties~\cite{PhysRevC.81.014303, Niksic2009Phys.Rev.C034303,Li2009Phys.Rev.C054301,Yang2023Phys.Rev.C024308}. We further confront the present approach with the isotonic shift between the charge radii of Sn and Cd isotopes.

\begin{figure}
	\centering
\includegraphics[width=0.9\linewidth]{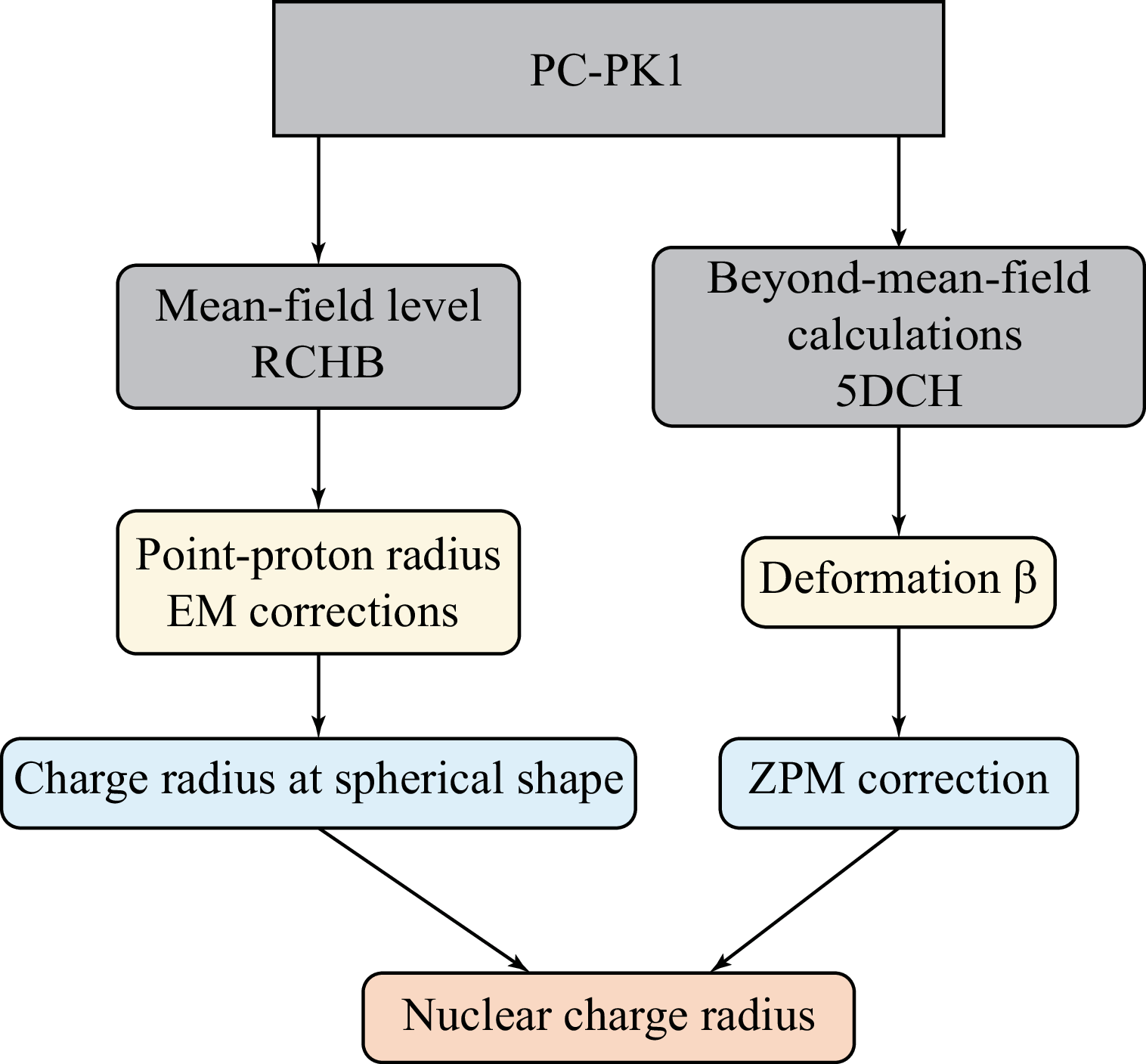}
\caption{(Color online) Schematic diagram of the self-consistent calculation of the mean-square charge radii $\langle r^2\rangle_\mathrm{ch}$ based on the relativistic density functional PC-PK1.
The relativistic continuum Hartree-Bogoliubov (RCHB) method is employed on the mean-field level, and a microscopically mapped five-dimensional collective Hamiltonian (5DCH) is used to take into account beyond-mean-field contributions.}\label{fig:fig0}
\end{figure}
According to Bohr's collective model~\cite{Bohr1975}, the mean-square charge radius of a nucleus reads
\begin{equation}\label{eq-rc}
  \langle r^2\rangle_{\rm ch}=\langle r^2\rangle_0\left(1+\frac{5}{4\pi}\beta^2\right)
\end{equation}
where $\langle r^2\rangle_0$ is the mean-square charge radius with no deformation, and $\beta$ is the quadrupole deformation parameter.
\textcolor{black}{Note that the quadrupole deformations for Ca isotopes are mainly correlated with the zero-point shape fluctuations. }
Although the above equation is somewhat semiclassical, it does not involve any parameter adjustment as in the previous works \cite{RING1996193, MENG2006470, Meng_2015, VRETENAR2005101, NIKSIC2011519,meng2013progress, XIA20181}, while reproducing the puzzling behavior of charge radii in the calcium isotopes. Therefore, the physics behind the result is rather clear—the beyond-mean-field effect related to shape vibration plays an important role in the charge radii of the calcium isotopes. In the future, it would be interesting to investigate this aspect using a more sophisticated, fully microscopic beyond-mean-field approach.
Throughout this work, the PC-PK1 density functional~\cite{PhysRevC.82.054319} is employed on the mean-field level and beyond to evaluate $\langle r^2\rangle_0$ and $\beta$, as illustrated in Fig. \ref{fig:fig0} and described as follows.

The spherical mean-square charge radius $\langle r^2\rangle_0$ is obtained by the spherical relativistic continuum Hartree-Bogoliubov (RCHB) method, which was developed based on the relativistic DFT with Bogoliubov transformation in the coordinate representation to provide a proper treatment of the pairing correlations and mean-field potentials in the presence of continuum~\cite{RING1996193, MENG2006470, Meng_2015, VRETENAR2005101, NIKSIC2011519,meng2013progress, XIA20181}.

Specifically, \textcolor{black}{the spherical mean-square charge radius} is calculated as
\begin{equation}\label{eq-r0}
  \langle r^2\rangle_0=\langle r^2\rangle_{\rm pp}+R^2_p+\langle r_p^2\rangle_{\rm so}+\frac NZ\left( R_n^2+\langle r_n^2\rangle_{\rm so}\right),
\end{equation}
where $\langle r^2\rangle_{\rm pp}$ denotes the mean-square point-proton radius, and the other terms originate from the intrinsic electromagnetic structure of nucleons. The proton~\cite{codata2018} and neutron~\cite{atac2021measurement} mean-square charge radii, $R^2_p=(0.8414\pm 0.0019)^2$ fm$^2$ and $R_n^2=-0.110\pm 0.008$ fm$^2$, account for the finite size of nucleons.
In addition, the proton and neutron spin-orbit charge radii $\langle r_{p,n}^2\rangle_{\rm so}$ also contribute to the charge radius as relativistic effects and could play an important role in describing the difference of charge radii between $^{40}$Ca and $^{48}$Ca~\cite{BERTOZZI1972408, PhysRevC.62.054303, PhysRevC.82.014320}.
In this work, $\langle r_{p,n}^2\rangle_{\rm so}$ are calculated from the single-particle wave function obtained in RCHB calculations~\cite{PhysRevA.107.042807, PhysRevC.110.064319}.

The quadrupole deformation $\beta$ in Eq.~(\ref{eq-rc}) is extracted from the quadrupole shape invariant following the Kumar-Cline sum rule~\cite{Kumar1972Phys.Rev.Lett.249, Cline1986Ann.Rev.Nucl.Part.Sci.683},
\begin{equation}\label{eq-beta2}
	\beta^2=\frac{16\pi^2}{25\left(eZ\right)^2\left(\langle r^2\rangle_0\right)^2}\sum_k B(E2;0_1^+\rightarrow 2_k^+).
\end{equation}
The collective ground state $0_1^+$ and excited $2^+$ states, as well as the \textcolor{black}{reduced electric quadrupole transition probabilities} $B(E2)$ values between them, are obtained by solving a microscopically-mapped five-dimensional collective Hamiltonian (5DCH) without additional free parameters~\cite{Niksic2009Phys.Rev.C034303,Li2009Phys.Rev.C054301}. 
In the calculation of $\sum_k B(E2;0_1^+\rightarrow 2_k^+)$ of Eq.~(\ref{eq-beta2}), the $30$ lowsest $2^+$ states are considered to obtain convergent results as shown in Ref.~\cite{Yang2023Phys.Rev.C024308}. 
These values, together with $B(E2)$ for the first $2^+$ state, i.e., $B(E2;0^+_1\to2^+_1)$, are compared with the experimental values~\cite{PRITYCHENKO20161} in Table \ref{tab-0}. 
The dominant contributions to the sum mostly come from the $2_1^+$ state, consistent with the prescription in Refs.~\cite {PhysRevC.28.355, PhysRevLett.59.2724}. $^{40}$Ca is somewhat the only exception. For the $B(E2;0_1^+\to 2_1^+)$ value, it is seen that the calculated value for $^{40}$Ca is underestimated, and those for $^{46,48,50}$Ca are overestimated. Similar overestimations were also found in the previous 5DCH calculations based on Gogny D1S interaction~\cite{PhysRevC.81.014303}. We suspect that such discrepancies are related to the fact that the 5DCH models do not take into account the noncollective degrees of freedom~\cite{PhysRevC.81.014303}, such as (multi-) quasiparticle excitations. 

Nevertheless, regardless of these discrepancies of $B(E2)$, the 5DCH model should be able to grasp the stiffness of the potential energy surface of the nucleus, thus adequately describing the quadrupole ZPM effects on the mean-square radii. 
It is important to note that the ZPM corrections presented here are derived from the shape invariants in Eq. (\ref{eq-beta2}), which differs from the approach used in previous 5DCH studies such as Ref. ~\cite{PhysRevC.81.014303}. In Ref.~\cite{PhysRevC.81.014303}, the shape distributions directly from 5DCH were used, i.e., the ZPM correction was obtained by directly averaging the charge radius over the full shape distribution in the $\beta$-$\gamma$ plane resulting from the 5DCH collective wave functions. 
It is found that the shape invariants and shape distributions lead to similar trend of the ZPM corrections along Ca isotopes.

The collective parameters that govern the dynamics of the Bohr Hamiltonian are fully determined by DFT calculations based on PC-PK1~\cite{Yang2021Phys.Rev.C054312,Yang2023Phys.Rev.C024308}.
It should be emphasized that $\beta^2$ could be significant even for a nucleus whose ground state is spherical on the mean-field level, e.g., Ca isotopes.
In such cases, the deformation is generated by the zero-point motions associated with the shape fluctuations around the spherical shape.


\begin{table}
	\caption{The calculated charge radii $R_\mathrm{ch}^\mathrm{Cal.}$ (fm) by Eqs.~(\ref{eq-rc}) and (\ref{eq-r0}), and the values of $\sum_k B(E2;0_1^+\rightarrow 2_k^+)$ $(e^2b^2)$ \textcolor{black}{together with $B(E2;0_1^+\to2_1^+)$ in square brackets} obtained by solving a microscopically-mapped 5DCH without additional free parameters~\cite{Niksic2009Phys.Rev.C034303,Li2009Phys.Rev.C054301}, in comparison with the experimental charge radii~($^{40-50}$Ca~\cite{ANGELI201369} and $^{52}$Ca~\cite{LI2021101440}) and $B(E2)\uparrow$ (i.e., $B(E2;0_1^+\to2_1^+)$~\cite{PRITYCHENKO20161}).}
	\centering
	\begin{tabular}{ccccc}
		\toprule
		Isotopes & $R_\mathrm{ch}^\mathrm{Cal.}$& $R_\mathrm{ch}^\mathrm{Exp.}$&$\sum_k B(E2;0_1^+\rightarrow 2_k^+)$ & $B(E2)\uparrow(\rm Exp.)$\\ \hline
		$^{40}$Ca&	$3.4893$&	$3.4776(19)$& $[0.0011]$ $0.0192$ & $0.00924(68)$\\
		$^{42}$Ca&	$3.5026$&	$3.5081(21)$& $[0.0384]$ $0.0436$ &$0.0369(20)$\\
		$^{44}$Ca&	$3.5136$&	$3.5179(21)$& $[0.0596]$ $0.0614$ & $0.0467(21)$\\
		$^{46}$Ca&	$3.5077$&	$3.4953(20)$& $[0.0547]$ $0.0556$ & $0.0168(13)$ \\
		$^{48}$Ca&	$3.4866$&	$3.4771(20)$& $[0.0267]$ $0.0303$ &$0.0092(^{+12}_{-5})$\\
        $^{50}$Ca&	$3.5046$&	$3.5168(26)$& $[0.0285]$ $0.0293$ &$0.00373(^{+20}_{-18})$\\
        $^{52}$Ca&	$3.5269$&	$3.5531(29)$& $[0.0294]$ $0.0332$ & \\
        \toprule
	\end{tabular}
    \label{tab-0}
\end{table}

\begin{figure}[!htbp]
	\centering
	\includegraphics[width=0.9\linewidth]{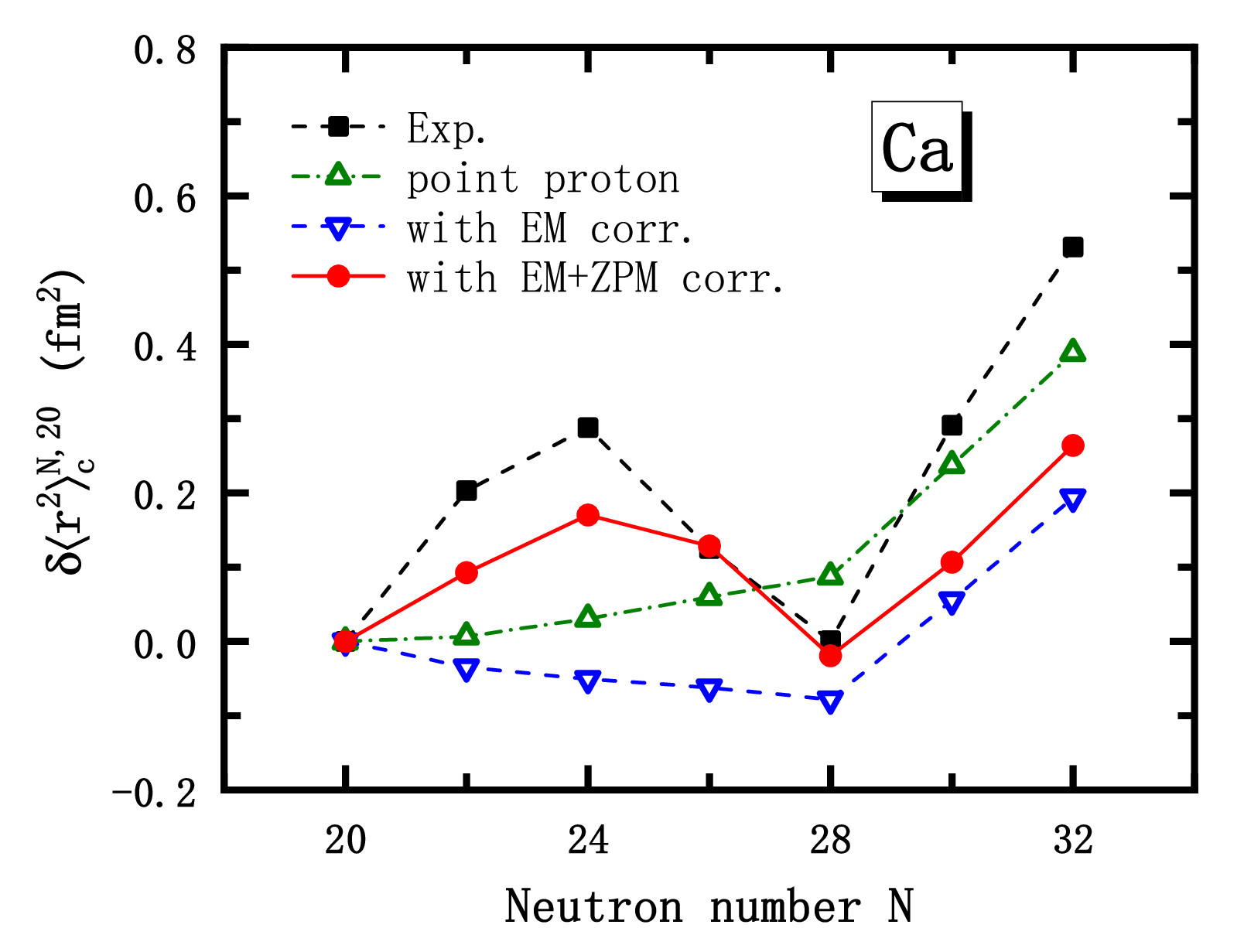}
  \caption{(Color online) The differential mean-square charge radius $\delta\langle r^2\rangle_c^{N,20}$ for the calcium isotopes.
  Squares indicate the experimental data~\cite{garcia2016unexpectedly,miller2019proton}.
  Upper triangles represent DFT results obtained by considering point-like protons, while lower triangles represent those including the intrinsic electromagnetic structure of protons and neutrons.
  The results further including the zero-point motion corrections, are shown by solid circles.
  }\label{fig:fig1}
\end{figure}

In Fig.~\ref{fig:fig1}, the trend of charge radii in Ca isotopic chain is shown by the differential mean-square charge radius, $\delta\langle r^2\rangle_{\rm ch}^{N,20}=\langle r^2\rangle_{\rm ch}^N-\langle r^2\rangle_{\rm ch}^{20}$, as a function of neutron number $N$.
From $N=20$ to $N=28$, the measured charge radii show a clear parabolic behavior with a maximum at $N=24$.
However, as is common in other DFT calculations \cite{RevModPhys.75.121, PhysRevC.84.054309, PhysRevC.93.034337, PhysRevC.99.034318, PhysRevC.104.064313, PhysRevC.105.034320}, the point-proton radii in the RCHB calculations with PC-PK1 increase monotonically as $N$ increases.
In contrast, the calculated results, including the intrinsic EM structure of nucleons, are monotonically decreasing.
Only by including both the intrinsic EM structure of nucleons and the zero-point motions in nuclear shape one can well reproduce the observed parabolic trend.
Moreover, the similar charge radii of $^{48}$Ca and $^{40}$Ca are also nicely predicted with $\delta\langle r^2\rangle_{\rm ch}^{28,20}=-0.019~\mathrm{fm}^2$, as compared to $\delta\langle r^2\rangle_{\rm ch}^{28,20}=0.088~\mathrm{fm}^2$ obtained using point-proton radii.
This is remarkable considering that the present calculations have no free parameters.
Note that the increasing trend of charge radii from $N=28$ to $N=32$ is also reproduced in the present calculations, but the magnitude is somewhat underestimated.

\begin{figure}[!htbp]
	\centering
	\includegraphics[width=0.9\linewidth]{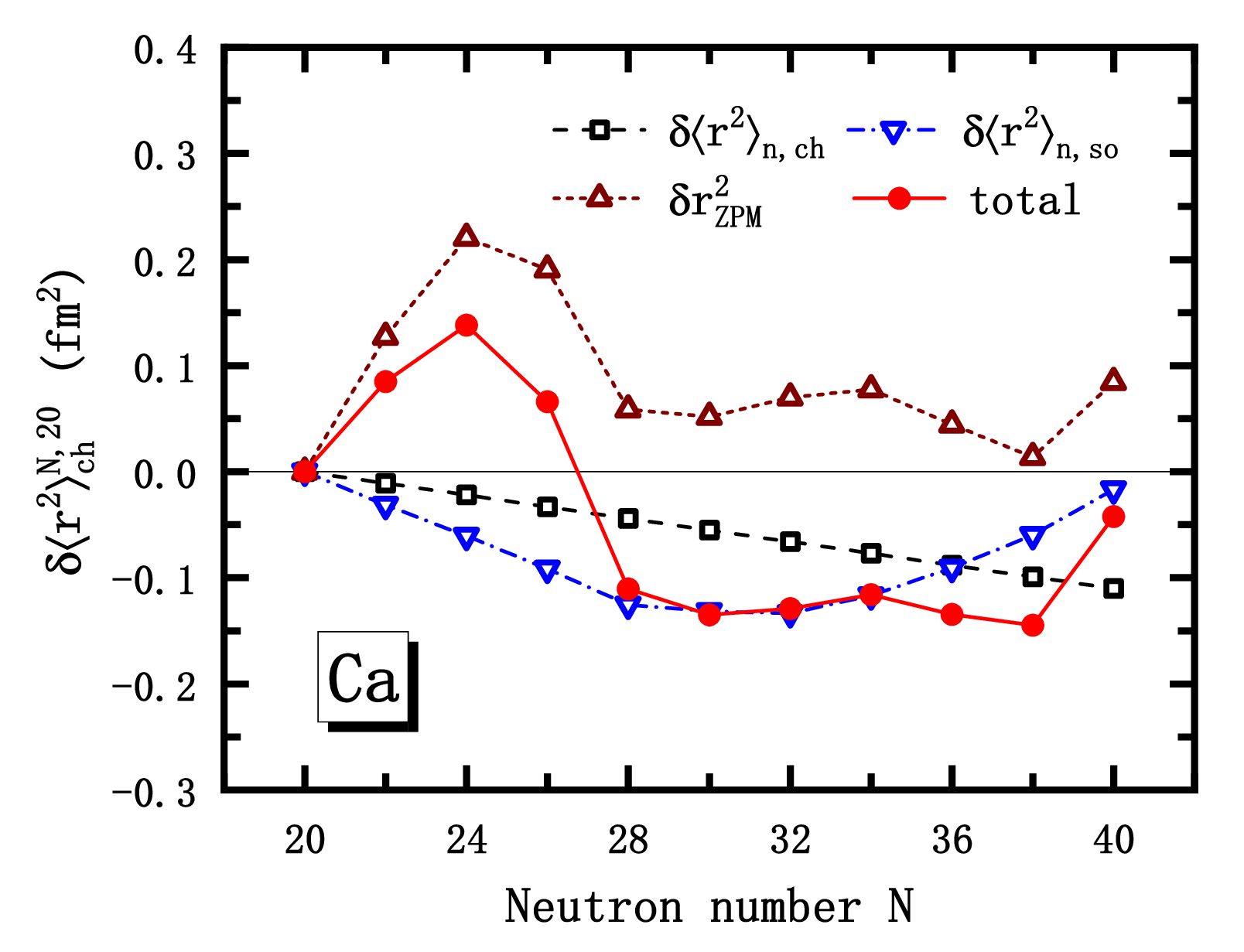}
	\caption{(Color online) The contributions from neutron charge radius (squares), neutron spin-orbit charge radius (lower triangle), and zero-point motions (upper triangle) to $\delta\langle r^2\rangle_\mathrm{ch}^{N,20}$ for Ca isotopic chain in the present calculations based on PC-PK1 density functional. See the text for details.}
	\label{fig:fig2}
\end{figure}

To understand the importance of the intrinsic EM structure of nucleons and the zero-point motions in nuclear shape, their contributions to $\delta\langle r^2\rangle_{\rm ch}^{N,20}$ are plotted in Fig.~\ref{fig:fig2}.
For the intrinsic EM structure of nucleons, the contributions from neutron charge radius $\delta\langle r^2\rangle_{\rm n,ch}\equiv\frac{\delta N}{Z}R_n^2$ and from neutron spin-orbit charge radius $\delta \langle r^2\rangle_{\rm n,so}\equiv\delta\left(\frac{ N}{Z}\langle r^2_n\rangle_\mathrm{so}\right)$ are separately displayed.
The proton counterparts contribute little and are therefore not shown.
With neutron number increases, the negative mean-square charge radius of neutron increasingly reduces $\delta\langle r^2\rangle_{\rm ch}^{N,20}$.
The neutron spin-orbit charge radius also provides a negative contribution, but the magnitude increases from $N=20$ to $28$ and then decreases from $N=28$ to $40$.
This mainly results from occupying the neutron $1f_{7/2}$ orbital that reduces the spin-orbit charge radius when moving from $N=20$ to $N=28$ and $1f_{5/2}$ orbital that has an opposite effect when moving from $N=28$ to $N=40$.
Note that the spin-orbit partners such as neutron $1f_{7/2}$ and $1f_{5/2}$ orbital contribute to spin-orbit charge radius with opposite sign~\cite{PhysRevC.62.054303}.
For the zero-point motion contributions $\delta r^2_{\rm ZPM}$, the parabolic behavior originates from the enhanced shape fluctuations around the spherical minimum for neutron-midshell Ca isotopes, because the potential energy surfaces of midshell Ca isotopes are softer than those of the closed-shell nuclei $^{40,48}$Ca~\cite{nuclearmap}.

\textcolor{black}{The comparison of calculated charge radii with the experimental values~\cite{ANGELI201369} for even $^{40-48}$Ca isotopes is shown in Table~\ref{tab-0}. The corresponding root-mean-square deviation for charge radii is \textcolor{black}{$0.0134$} fm, a little higher than the experimental error.
}

\begin{figure}[!htbp]
	\centering
	\includegraphics[width=1.0\linewidth]{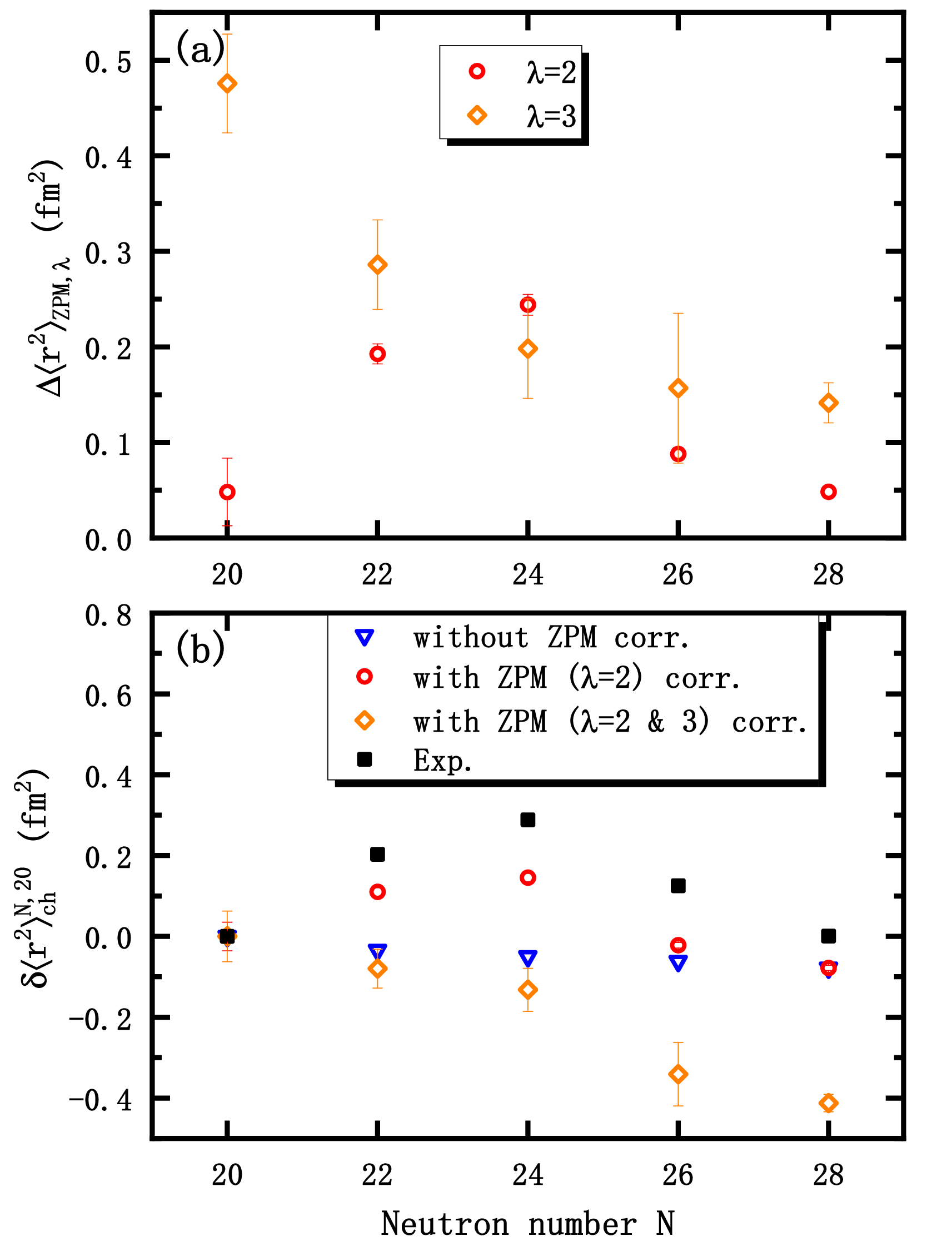}
	\caption{(Color online) (a) The contributions of ZPM from quadrupole and octupole vibrations to mean-square charge radius for Ca isotopic chain. (b) The differential mean-square charge radius with and without ZPM corrections for Ca isotopes. Where the recommended experimental $B(E\lambda)$ values listed in Tables~\ref{tab-0} and~\ref{tab-1} are used. 
    }
\label{fig:fig3}
\end{figure}

On the other hand, it has been demonstrated in Refs.~\cite{BARRANCO198590, PhysRevLett.59.2724, PhysRevC.28.355, FAYANS200049} that the contribution of ZPM from the octupole vibration visibly impacts the results for Ca isotopes in addition to the quadrupole vibrations. 
On the basis of the adopted experimental $B(E2;0_1^+\to2_1^+)$ and $B(E3;0_1^+\to3_1^-)$ values shown in Tables~\ref{tab-0} and~\ref{tab-1}, the contributions from quadrupole and octupole vibrations are evaluated by the following expressions~\cite{FAYANS200049}, as shown in Fig.~\ref{fig:fig3}.
	\begin{equation}
		\Delta\langle r^2\rangle_{\mathrm{ZPM},\lambda}=\frac{5}{4\pi}\langle r^2\rangle_0\langle\beta_\lambda^2\rangle,
	\end{equation}
with $\langle \beta_\lambda^2\rangle=\left(\frac{4\pi}{3ZeR_0^\lambda}\right)^2B(E\lambda;0_1^+\to f)$, where $f$ denotes $2_1^+$ for $\lambda=2$ and $3_1^-$ for $\lambda=3$, respectively, and $R_0^2=\frac53\langle r^2\rangle_0$.   
It can be seen from Fig.~\ref{fig:fig3}(a) that the contributions for octupole vibration $\Delta\langle r^2\rangle_{\mathrm{ZPM},\lambda=3}$ are monotonously decreasing along the Ca isotopes, which almost consistent with the results proposed by Barranco and Broglia~\cite{BARRANCO198590}. 
Moreover, as shown in Fig.~\ref{fig:fig3}(b), the results with both quadrupole and octupole ZPM contributions get worse on account of the decreasing trend of the latter. 
According to the results of Ref.~\cite{BARRANCO198590}, even if the contributions associated with the other multipolarities (such as $\lambda=0,4,5$) are taken into account, the description of charge radii for Ca isotopes would be hardly improved. 
Therefore, their accurate description might require re-evaluation of the coupling parameters in the density functional when including higher multipolarity ZPM effects.

\begin{table}
	\caption{The recommended experimental values ($^{40-46}$Ca:~\cite{KIBEDI200235}; $^{48}$Ca:~\cite{PhysRevC.83.044327}) of $B(E3;0_1^+\to 3_1^-)$ in unit of $e^2\mathrm{b}^3$.
    }	\centering
	\setlength{\tabcolsep}{4mm}{
	\begin{tabular}{cc}
		\toprule
		Isotopes & Adopted \\ \hline
		$^{40}$Ca&	$0.0184(20)$\\
		$^{42}$Ca&	$0.0110(18)$\\
		$^{44}$Ca&	$0.0076(20)$\\
		$^{46}$Ca&	$0.006(3)$  \\
		$^{48}$Ca&	$0.0054(8)$ \\\toprule
	\end{tabular}}\label{tab-1}
\end{table}

\begin{figure}[!htbp]
	\centering
	\includegraphics[width=1.0\linewidth]{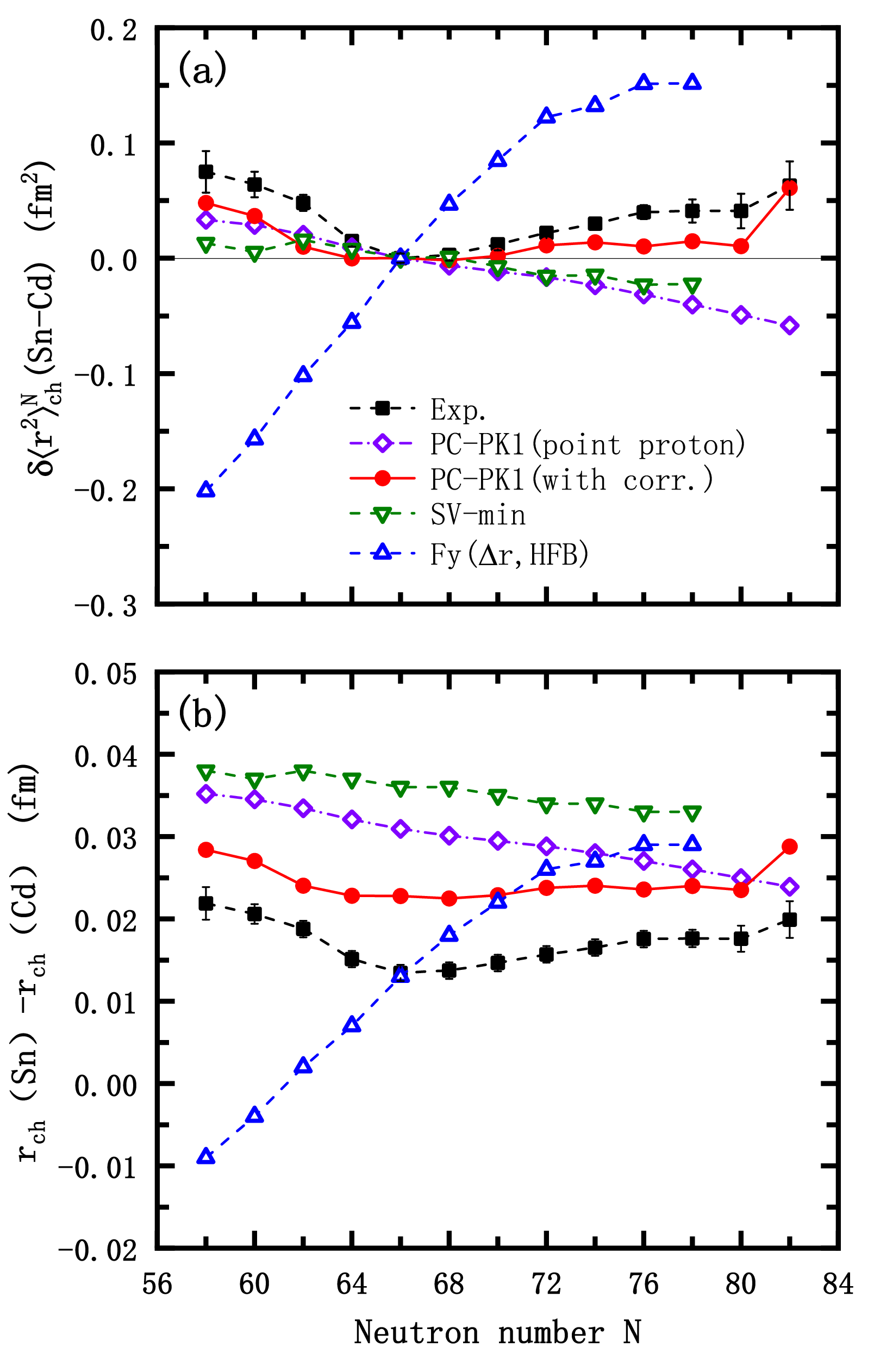}
	\caption{(Color online) 
		The isotonic shifts between the Sn and Cd isotopes predicted by present calculations with PC-PK1, as functions of neutron number.
		In panel (a), the trend of isotonic shifts $\delta \langle r^2\rangle^N_{\rm ch}$(Sn-Cd)=$\langle r^2\rangle^N_{\rm ch}$(Sn)-$\langle r^2\rangle^N_{\rm ch}$(Cd) is shown, where the results are the relative difference with respect to $\delta \langle r^2\rangle^{N=66}_{\rm ch}$(Sn-Cd).
		In panel (b), the isotonic differences of charge radii between Sn and Cd are displayed.
		The experimental data are taken from Refs.~\cite {PhysRevLett.121.102501, PhysRevLett.122.192502}.
		In addition, the results of spherical HFB calculations with SV-min and Fy($\Delta r$, HFB) density functional taken from Refs.~\cite {PhysRevLett.121.102501, PhysRevLett.122.192502} are also shown for comparison.
	}
\label{fig:fig4}
\end{figure}

Furthermore, the predictive power of present calculations is also examined in the isotonic shift between the charge radii of Sn and Cd isotopes, $\delta \langle r^2\rangle^N_{\rm ch}$(Sn-Cd)=$\langle r^2\rangle^N_{\rm ch}$(Sn)-$\langle r^2\rangle^N_{\rm ch}$(Cd).
In Fig.~\ref{fig:fig4}(a), the trend of Sn-Cd isotonic shifts is shown for present spherical RCHB calculations with PC-PK1 density functional, in comparison with the experimental result~\cite{PhysRevLett.122.192502, PhysRevLett.128.022502}.
The isotonic shifts extracted from the point-proton radii with PC-PK1 monotonically decrease with neutron number.
With the EM and ZPM contributions included self-consistently, the experimentally observed trend with a minimum at $N=66$ is overall reproduced.

The results of the previous spherical Hartree-Fock-Bogoliubov (HFB) calculations with SV-min~\cite{PhysRevC.79.034310} and Fy($\Delta r$, HFB)~\cite{miller2019proton} density functional \textcolor{black}{given in Refs.} ~\cite{PhysRevLett.121.102501,PhysRevLett.122.192502} are also shown in Fig.~\ref{fig:fig4}(a) for comparison.
The SV-min provides a trend similar to PC-PK1 when only point-proton contributions are considered.
In Ref.~\cite{PhysRevC.106.L011304}, it was shown that the inclusion of ZPM corrections using the experimental $B(E2)$ improves the SV-min results.
In the future, it would be interesting to investigate whether such an improvement still holds when including both the EM and ZPM contributions self-consistently, as in the present PC-PK1 calculations.\
In stark contrast to PC-PK1 and SV-min, Fy($\Delta r$, HFB) predicts a drastically increasing isotonic shift, which is in poor agreement with experimental data.
As shown in Ref.~\cite{PhysRevLett.122.192502}, Fy($\Delta r$, HFB) can well reproduce the kink at $^{132}$Sn and $^{208}$Pb, while it produces an underestimation of charge radii of Sn isotopes for $N < 64$ region and an overestimation for $64 <N <82$ region, which lead to such poor agreement.
For Fy($\Delta r$, HFB), taking into account the EM and ZPM contributions is not an option.
Because they will spoil the description of the parabolic behavior of charge radii in Ca isotopes, which is delicately adjusted in the fitting procedure of Fy($\Delta r$, HFB) without the EM and ZPM contributions~\cite{PhysRevC.95.064328}.

Apart from the trend of Sn-Cd isotonic shifts, the absolute value of Sn-Cd isotonic shift is also improved by the inclusion of EM and ZPM contribution, as shown in Fig.~\ref{fig:fig4}(b).
The present PC-PK1 results are in better agreement with the experimental data compared to the results of SV-min and Fy($\Delta r$, HFB).
Nevertheless, a slight overall overestimation remains due to the somewhat underestimation of charge radii for the Cd isotopic chain.

In summary, the parabolic part of the observed anomaly in the charge radii along the Ca isotopes is reproduced, for the first time, in a self-consistent approach free of local parameter adjustments.
Particular emphasis is placed on the important role of the intrinsic electromagnetic structure of nucleons and the zero-point motions of nuclear shape in nuclear charge radii, which have often been neglected in existing DFT calculations.
Based on the relativistic density functional PC-PK1, the former and latter contributions are taken into account, respectively, with the relativistic continuum Hartree-Bogoliubov method on the mean-field level and a microscopically mapped Bohr Hamiltonian beyond the mean-field level.
The intrinsic neutron (spin-orbit) charge radius provides a significant decreasing trend from $^{40}$Ca to $^{48}$Ca, crucial for the explanation of the similar charge radii between these two nuclei.
The zero-point motions are responsible for the inverse parabolic behavior, due to the enhanced shape fluctuations around the spherical minimum for neutron-midshell Ca isotopes compared to the closed-shell $^{40,48}$Ca.
To accurately reproduce the linear dependence of charge radii at the mean-field level while accommodating ZPM corrections from both quadrupole and octupole vibrations, one could refine the coupling parameters. This refinement should simultaneously account for both intrinsic electromagnetic structure and ZPM effects from both quadrupole and octupole vibrations during the parameter optimization process.

The present approach is also confronted with the isotonic shift between the charge radii of Sn and Cd isotopes and found to significantly improve the agreement with experimental data.
Therefore, the present work is expected to impact the description of nuclear charge radii throughout the nuclear landscape.

The work of H. H. X. and J. L. was supported by the National Natural Science Foundation of China (Grants No. 12475119, No. 11675063) and the Natural Science Foundation of Jilin Province (No. 20220101017JC).
The work of Y. L. Y. and P. W. Z. has been supported in part by the National Natural Science Foundation of China (Grants No. 12141501, No. 12070131001, No. 11935003, No. 11975031), and the High-performance Computing Platform of Peking University.
Y. L. Y. and P. W. Z. acknowledge the funding support from the State Key Laboratory of Nuclear Physics and Technology, Peking University.

\bibliography{apssamp}

\end{document}